\documentclass{sig-alternate}

\usepackage{amsmath}
\usepackage{amssymb}

\usepackage{color}
\usepackage{dcolumn}
\usepackage{float}
\usepackage{graphicx}
\usepackage[latin9]{inputenc}
\usepackage{multirow}
\usepackage{rotating}
\usepackage{subfigure} 
\usepackage{psfrag}
\usepackage{tabularx}
\usepackage[hyphens]{url}
\usepackage{wrapfig}
\usepackage{fancyvrb}

\usepackage[bookmarks, bookmarksopen, bookmarksnumbered]{hyperref}
\usepackage[all]{hypcap}
\newcommand{\rcltool}{\texttt{GetComponents}}

\usepackage[top=2.5cm, bottom=2.5cm, left=2.5cm, right=2.5cm]{geometry}

\sloppypar

\begin{document}

\title{Simplifying Complex Software Assembly: The Component Retrieval Language and Implementation}

\conferenceinfo{TeraGrid}{`10, August 2-5, 2010, Pittsburgh, PA, USA.}
\CopyrightYear{2010}
\crdata{978-1-60558-818-6/10/08}

\numberofauthors{5}

\author{
\alignauthor
Eric L. Seidel \\
	\affaddr{City College of New York} \\
	\affaddr{New York, NY 10031} \\
	\affaddr{(Presenting author)} \\
\alignauthor
Gabrielle Allen\\
	\affaddr{Department of Computer Science} \\
	\affaddr{Center for Computation \& Technology} \\	
	\affaddr{Louisiana State University} \\
	\affaddr{Baton Rouge, LA 70803} \\
\alignauthor
Steven Brandt \\
	\affaddr{Center for Computation \& Technology} \\
	\affaddr{Louisiana State University} \\
	\affaddr{Baton Rouge, LA 70803} \\
\and
\alignauthor
Frank L\"offler\\
	\affaddr{Center for Computation \& Technology} \\
	\affaddr{Louisiana State University} \\
	\affaddr{Baton Rouge, LA 70803} \\
\alignauthor
Erik Schnetter\\
	\affaddr{Department of Physics \& Astronomy} \\
		\affaddr{Center for Computation \& Technology} \\	
	\affaddr{Louisiana State University} \\
	\affaddr{Baton Rouge, LA 70803} \\
}




  
    
\maketitle
\begin{abstract}
Assembling simulation software along with the associated tools and utilities is a challenging endeavor, particularly when the components are distributed across multiple source code versioning systems. It is problematic for researchers compiling and running the software across many different supercomputers, as well as for novices in a field who are often presented with a bewildering list of software to collect and install. 

In this paper, we describe a language (CRL) for specifying software
components with the details needed to obtain them from source code
repositories. The language supports public and private access.   We
describe a tool called \rcltool\ which implements CRL and 
can be used to assemble software. 

We demonstrate the tool for application scenarios with the Cactus
Framework on the NSF TeraGrid resources. The tool itself is
distributed with an open source license and freely available from our
web page.
\end{abstract}

\maketitle



\section{Introduction}

Compute resources, along with their associated data storage and network connectivity, are growing ever more powerful. 
The current computational environment provided by the National Science Foundation to support its academic research agenda 
includes several petascale machines as part of the distributed TeraGrid facility and the  multi-petaflop  ``Blue Waters'' machine which should
 be operational in 2011. 
This increase in compute capacity is needed to satisfy the requirements of software applications that are being developed to model Grand Challenge 
scientific problems with unprecedented fidelity in fields such as climate change, nuclear fusion, astrophysics, material science as well as non-traditional applications in social sciences and humanities.
As these applications grow in size they are also growing more complex; coupling 
together different physical models across varying spatial and temporal scales, and involving distributed teams of interdisciplinary researchers, 
heralding a new era of collaborative multi-scale 
and multi-model simulation codes. 

\begin{figure*}[ht!]
\centerline{\includegraphics[width=\linewidth]{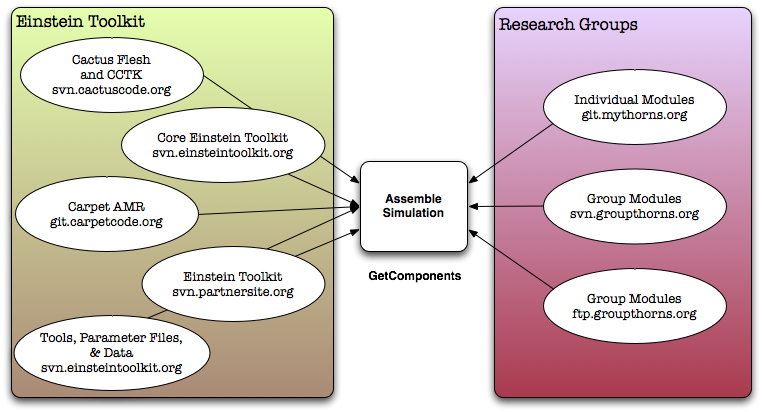}}
\caption{\label{etexample} Applications such as the Einstein Toolkit (Section~\ref{einstein}) built from component frameworks such as Cactus can involve assembling hundreds of modules from distributed, heterogeneous source code repositories.}
\end{figure*}

One approach to developing application codes in an efficient,  sustainable and extensible manner is through the use of application-level component frameworks or 
programming environments. Component frameworks can support  reuse and community development of software 
by encapsulating common tools or methods within a domain or set of domains. 
Cactus, a component framework for high performance computing~\cite{Cactusweb,Goodale02a}, provided the motivation for the work described in this paper. 
As we describe in Section~\ref{cactus}, Cactus users typically assemble their simulation codes from many different 
software modules distributed from different locations, providing a number of challenges for users in both describing the needed modules and actually 
retrieving them (Figure~\ref{etexample}).

Version control systems, such as CVS, Subversion, or Git, are used to manage and maintain the modules that make up these frameworks. Such systems track changes to the source code and allow developers to recover a stable version of their software, should an error be introduced. There are a large number of version control systems in use, and while some are relatively compatible (tools exist to convert a CVS repository to Subversion or Git), many are not.\footnote{An in-depth comparison of version control system can be found at \url{http://en.wikipedia.org/wiki/Comparison_of_revision_control_software}} This can create issues when users want to assemble, and then maintain, a component framework that includes modules from a variety of systems. A complex framework like Cactus would be very difficult to maintain without some way of automating the checkout/update process.

To address this issue in a general manner for complex code assembly
for any application, we have designed a new language, the Component
Retrieval Language (or CRL) that can be used to describe modules along
with information needed for their retrieval from remote, centralized
repositories.  We have implemented a tool based on this language that
is now being used by Cactus users for large scale code assembly. 

This paper starts by describing the Cactus framework~\ref{cactus}, which provides the motivation for the Component Retrieval Language. Then it describes related work in Section~\ref{related} before detailing the design issues for the component retrieval language in 
Section~\ref{design}. Section~\ref{crl} describes the grammar of the new component retrieval language, and Section~\ref{gc} discusses the 
\rcltool\ tool that has been written to implement this language. Section~\ref{examples} provides an example showcasing the
 use of \rcltool\ on the resources of the NSF TeraGrid for a community of Cactus users. Section~\ref{future} describes planned future work in improving
  code assembly for complex software efforts in scientific computing before concluding in Section~\ref{conclusions}.

\section{Cactus Example for Distributed Code Assembly}
\label{cactus}

Cactus~\cite{Cactusweb,Goodale02a} is an open-source framework designed for the collaborative development of large scale simulation codes in science and engineering. 
Computational toolkits distributed with Cactus already provide a broad range of capabilities for solving initial value problems in a parallel environment. 
The Cactus Computational Toolkit includes modules for I/O, setting up coordinate systems, outer and symmetry boundary conditions, domain decomposition and message passing, standard reduction and interpolation operators, numerical methods such as method of lines, as well as tools for debugging, remote steering, and profiling.
Cactus is supported and used on all the major NSF TeraGrid machines, as well as others outside the TeraGrid, and is included in the advanced tools development for the NSF Blue Waters facility.

Cactus is used by applications in areas including relativistic
astrophysics, computational fluid dynamics, reservoir simulations,
quantum gravity, coastal science and computer science.
Cactus users assemble their codes from a variety of independent components (called \emph{thorns}) which are typically developed
 and distributed from different source code repositories which are geographically, institutionally and politically varied.  
 Source code repositories can be public (with anonymous read access), private (with authentication by user or group); 
 they are of different types (e.g.\ CVS, SVN, darcs, git, Mercurial); and the location of thorns within a repository varies. Cactus simulations, 
for example in the field of numerical relativity, can involve some
200 thorns from some ten different repository servers around the world. 

In addition, Cactus users typically use other tools or utilities that are not part of the actual simulation code, such as the 
Simulation Factory~\cite{simfactory} for building and deploying, or visualization clients and shared parameter files.

\section{Related Work}
\label{related}

Cactus already included a tool for assembling codes from thorns;
\texttt{GetCactus} that was released in 1999 with the first
general release of Cactus and addressed several of the issues alluded
to in the introduction.  \texttt{GetCactus} was written specifically
to check out Cactus thorns, with a rudimentary syntax~\cite{getcactus}
that built on the existing concept of a Cactus \emph{thorn list}.
When \texttt{GetCactus} was designed and implemented, in addition to
being specific for Cactus, it only supported the use of CVS
repositories.  One issue that has become more serious as thorn lists
have become longer is the difficulty in distributing thorn lists to
others, since editing of the thorn list is required to change
authentication details that are user specific.


A rudimentary syntax for code assembly is also provided by the
\emph{NMI} Build and Test Lab~\cite{nmibuildandtest} that provides
infrastructure for automated downloading, building, and testing of
complex applications and software infrastructures on a set of commonly
used architectures.  NMI provides access to actual hardware on which
the test is run, focuses on reproducible results by providing
well-defined test systems, and offers a web-based user interface to
browse and examine test results.  To download codes for testing, NMI
supports CVS and SVN repositories directly, but only simple scenarios
are supported, and each component's location has to be described in a
separate file.  In addition, NMI supports a generic \emph{fetch} stage
where a user-defined script can execute arbitrary code to download
components.  To build and test Cactus at NMI, we first download
\rcltool\ and a thorn list via SVN, and then run \rcltool\ to retrieve
Cactus and the desired components.

\emph{ETICS} (eInfrastructure for Testing, Integration and
Configuration of Software) and its successor ETICS~2 are similar to
NMI\@.  They focus on dependencies between packages, testing, and
reproducibility and certification of results.  That is, the emphasis
is on testing a snapshot of project in very well defined environment
(e.g.\ ``test on MacOS X 10.4 with a 32-bit PowerPC processor, kernel
version 8.8.0, and using gcc 4.0.1'').  This addresses the needs of
integrators and managers, who can assume that a project is releasing
software in a shrink-wrapped manner.  \rcltool, on the other hand,
addresses the needs of software developers that need to handle and
assemble components long before the shrink-wrap stage of a project has
been reached.  (In fact, in a research environment, software is often
never publicly released since the potential user base is too small;
instead, it is only informally shared among colleagues.)

\emph{BuildBot}~\cite{buildbotweb} is a Python-based system to
automate building and testing.  It is much simpler than NMI or ETICS,
and consists only of software that the user installs, without
providing actual testing hardware.  Being Python based, software is
checked out via commands in a Python script.  BuildBot provides some
abstraction to access CVS, SVN, etc.\ repositories, but the download
process is described in a procedural manner as sequence of commands,
not in a descriptive manner.  This means that the information that has
been specified to download the software is ``hidden'' in the Python
script and is not accessible to other tools.

Debian, Red Hat, SUSE etc.\ are Linux distributions where a complete
installation consists of a set of \emph{packages}.  These packages are
available in a specific format (e.g.\ \texttt{deb}, \texttt{rpm})
which contains their source code (or binary code) as well as metadata
describing e.g.\ package dependencies and installation procedures.
Usually, these packages are available from a single, centralized
source (e.g.\ the distributor itself), and they thus do not need to
address the issues that \rcltool\ addresses.

Ubiqis uses a naming system where components are completely identified
(including their location and version) by means of a uniquely constructed
package name~\cite{brandt2008ubiqis}.
References to this package can be automatically detected
and downloaded in response to file system access using FUSE~\cite{FUSE}.
In principle, this means that versioning information can automatically
be sorted out for any component distribution. In cases where dependencies
are not automatic, it makes it possible to search the community
component space in a convenient way. However, it requires that its referents
be immutable, and it downloads packages from the
web instead of communicating with source code control systems. While it
addresses somewhat different issues than the current work, there is
synergy and the possibility exists to make use of some Ubiqis or some
variant of it in the future.

\section{Design Issues}
\label{design}

Based on our experiences with the Cactus Framework and its different user communities, we identified the following needs for the component retrieval language:

\begin{itemize}

\item \emph{Easy distribution of component lists.} The component lists (or CRL files) should be able to be constructed such that they can be distributed and used without editing. For Cactus users, this had been a growing issue with the \texttt{GetCactus} format where each entry in files would  typically need to be edited to change the username for each repository. The ability to easily distribute and pass  on 
Cactus thorn lists is a crucial step in simplifying Cactus for new users.

\item \emph{Support for both anonymous and authenticated retrieval of components.} Authenticated checkout of components is important for developers that will be committing changes back to a software repository, or for software that is restricted in access. Such a capability
 is important for the Cactus community where many users are also developing components. Authentication is handled differently by different versioning systems (for example, CVS requires an "anonymous username/password" for users to perform an anonymous checkout, whereas Subversion and Git do not), further users can have different authentications for different systems (e.g.\ different usernames and passwords).  

\item \emph{Support for different repository and distribution types:} Cactus thorns across the community are currently distributed from CVS, Subversion 
(SVN) and git source code repositories, with Mercurial being a likely choice in the future. 
Other common distribution mechanisms for software components include Darcs and simple HTTP/FTP downloads. 
  
\end{itemize}

In addition to supporting these features of the CRL, the implemented tool should:

\begin{itemize}  
  
\item \emph{Support updating components:} Source code repositories using CVS, SVN, etc, support updating of software, making it possible
 for developers to merge changes from others with their own code changes. The  retrieval tool should handle updates 
 in a manner suitable for developers.  

\item \emph{Handle multiple component lists:} This allows a community to share a common component list, which can be extended via additional component lists for a research group and/or individual.

\item \emph{Handle distributed version control systems:} The nature of distributed systems such as git or mercurial require that one 'clone' an entire repository instead of retrieving individual components. This is inconvenient when trying assemble complex software frameworks that only require a few components from a distributed repository. The retrieval tool should be able to process an entire repository while presenting only the components that have been requested by the user.

\end{itemize}

\begin{table}
\centering
\caption{\label{directives1} CRL directives (outside component block)}
\begin{tabular}{|l|p{50 mm}|}
\hline
\textbf{CRL Directive} & \textbf{Description} \\ \hline \hline
\texttt{CRL\_VERSION} & Currently 1.0. Also indicates that the file is in CRL format, so it must be the first non-comment line in each component list. \\ \hline
\texttt{DEFINE} &  User-defined terms that will be reflected throughout the rest of the component list. \\
\hline
\end{tabular}
\end{table}%
\begin{table}
\centering
\caption{\label{directives2} CRL directives (inside component block)}
\begin{tabular}{|l|p{50 mm}|}
\hline
\textbf{CRL Directive} & \textbf{Description} \\ \hline \hline
\texttt{TARGET} &  Placing of component relative to the current directory. \\ \hline
\texttt{TYPE} &  Tool used to checkout the component.  \\ \hline
\texttt{URL} &  Repository location. \\ \hline
\texttt{AUTH\_URL} &  Repository location for authenticated access. Only needs to be set if the URL for authenticated access is different from the URL for anonymous access. \\ \hline
\texttt{ANON\_USER} &  Username associated with an anonymous checkout. \\ \hline
\texttt{ANON\_PASS} &  Password associated with an anonymous checkout. \texttt{!ANON\_PASS} must be set if \texttt{!ANON\_USER} is. \\ \hline
\texttt{REPO\_PATH} &  Prefix for retrieving components from a git or mercurial repository, when a directory structure different than that provided by the repository is needed. \\ \hline
\texttt{CHECKOUT} &  Components to be retrieved from a repository. Multiple components are separated by one or more newlines. \\ \hline
\texttt{NAME} &  Alternate name for checkout directory if required. \\ \hline
\end{tabular}
\end{table}%

\section{The Component Retrieval Language} 
\label{crl}

This section provides a formal description of the Component Retrieval Language (CRL).  
In designing the CRL we did not seek to replicate all possible 
features of existing version control systems, but to encapsulate the functionality required by our 
considered use cases and allow for future extensibility. Further, a careful distinction was kept between
 the underlying language and implementation specific details in the \rcltool\ tool. 

The resulting Component Retrieval Language  has eleven different directives which are described in Tables~\ref{directives1} and \ref{directives2}. 
Files written using the CRL are structured 
with a header section that defines the version of the CRL in the file, which serves the dual purpose of identifying the file as a CRL list, and providing a way to determine compatibility with future updates of the language. It also sets up user defined variables that can simplify maintaining a long component list. Following the
 header section, the rest of the file consists of 
\emph{component blocks}, with each block of components having a common repository description.

The Component Retrieval Language also has an associated grammar written in Bison~\cite{bisonweb}
 (a variant of the Backus-Naur Form (BNF)), which is shown in Figure~\ref{bnf}. 
 While the grammar is fairly simple, it is nonetheless useful to provide a formal specification. This provides assurance
that the grammar is unambiguous, and provides a complete and succinct (albeit somewhat mathematical) form of documentation for the syntax.

\begin{figure}[t!]
{\small
\begin{Verbatim}[frame=single, framerule=0.3mm]
# NAME is an alphanumeric or '.' character

DOCUMENT : DIRECTIVES ;
        
DIRECTIVE : DEFINE NAME '=' PATH EOL           
          | CHECKOUT '=' COMPONENTLIST EOL     
          | CHECKOUT '=' EOL COMPONENTLIST EOL 
          | REPO_LOC '=' LOC EOL               
          | AUTH_LOC '=' LOC EOL               
          | PATH_DIRECTIVE '=' PATH EOL 
              # !REPO_PATH, !CHECKOUT, !TARGET,
              # !ANON_PASS, !NAME
          | NAME_DIRECTIVE '=' NAME EOL 
              # !CRL_VERSION, !AUTH_USER,
              # !ANON_USER, !TYPE
          ;

DIRECTIVES : DIRECTIVE
           | DIRECTIVES DIRECTIVE
           ;
             
LOC : PSERVER PATH            # CVS repository
    | NAME ':' '/' '/' PATH   # Git/SVN repository
    | NAME '@' NAME ':' PATH  # Git repository
    ;

PATH : NAME                         
     | '/' NAME                  
     | PATH '/' NAME              
     ;

COMPONENTLIST : PATH 
              | COMPONENTLIST EOL PATH ;
\end{Verbatim}
}
\caption{\label{bnf}Grammar for the CRL in Bison format}
\end{figure}

\section{GetComponents: A CRL Implementation}
\label{gc}

This section describes a Perl script, called \rcltool, which was developed to process CRL files and retrieve the indicated components. Perl was chosen because it is quick,
 lightweight, and it has a very powerful regular expression engine to parse the component lists. \rcltool\ can currently retrieve components from CVS, Subversion, 
 Git, Darcs and Mercurial repositories, as well as via http and ftp downloads. It provides multiple command line options as seen in Table~\ref{cmdoptions}, including anonymous mode, automatic
  updates, two levels of verbosity, and overriding the root directory for the components. Anonymous mode will force all checkouts to use anonymous methods. The auto-update option will bypass the user prompt and update any components
   that have been previously checked out, this allows \rcltool\ to be safely called by another program as a background process.

\begin{table*}
\centering
\begin{tabular}{|l|p{115mm}|}
\hline
\textbf{Command-line Option} & \textbf{Description} \\ \hline \hline
\texttt{--help} & Print a brief help message and exit. \\ \hline
\texttt{--man} & Print the full man page and exit. \\ \hline
\texttt{--verbose} & Print all system commands as they are executed by script. A second level of verbosity, declared by -v -v, will also display the output from the system commands. \\ \hline
\texttt{--debug} & Print a list of components that will be checked out or updated, along with the total number of components in the list. \\ \hline
\texttt{--anonymous} & Override any stored login credentials and use anonymous checkouts for all components. \\ \hline
\texttt{--update} & Override the update prompt and process all updates. \\ \hline
\texttt{--root} & Override the root directory in the component list. This allows checking out into an arbitrary directory. \\ \hline
\texttt{--reset-authentication} & Delete any CRL authentication files before processing the component list. \\ \hline
\end{tabular}
\caption{\label{cmdoptions}The command-line options for \rcltool.}
\end{table*}

\begin{figure*}[t]
\centerline{\includegraphics[width=\linewidth]{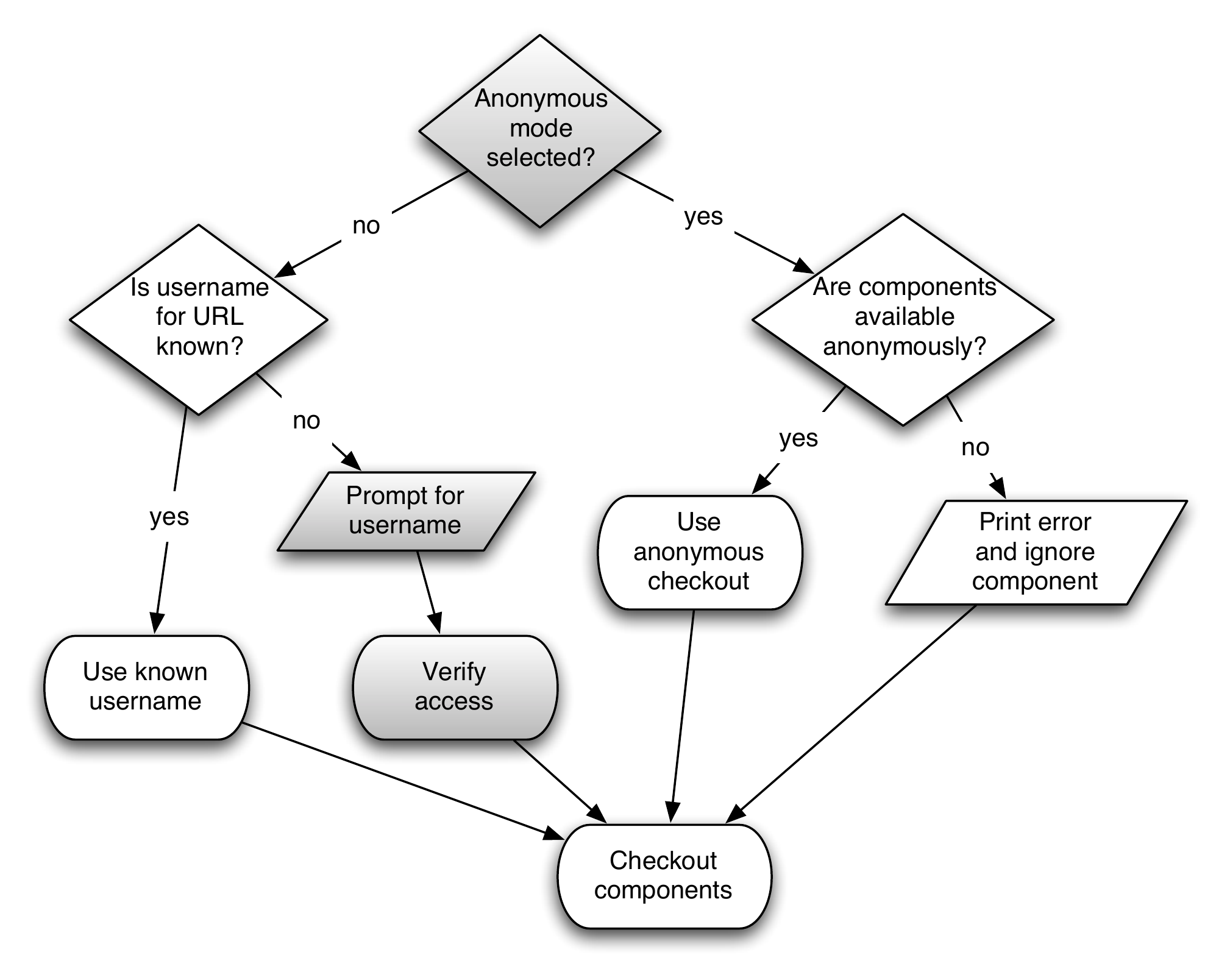}}
\caption{\label{authentication} Process for authentication implemented in the GetComponents tool. Authentication is defined on component blocks. Shaded areas indicate user interaction.}
\end{figure*}

Authentication and updates are handled by the underlying version control tools, with \rcltool\ providing a uniform layer between the user and the underlying tools. Figure~\ref{authentication} shows the general authentication process used by \rcltool, which is called once for each component block, unless anonymous mode has been selected. It first checks for {\tt !AUTH\_URL}, which specifies authenticated access to the repository. It then attempts to match the  {\tt AUTH\_URL} to the \rcltool\ users file (located by default in {\tt \$HOME/.crl/users}). If a match is found, \rcltool\ will use the associated username and then proceed to processing the next component block. If no match is found, \rcltool\ will prompt the user for their username, and attempt to login to the repository using the appropriate command (eg. {\tt cvs login}), after which it will save the username and URL in the users file. This has the security benefit of keeping passwords visible only to the actual retrieval tools. The user may also specify a '-' at this prompt to indicate they wish to perform an anonymous checkout for all components in the block. \rcltool\ will store this as well in the users file, so the user is not forced to specify anonymous access repeatedly. If the user mistakenly entered the wrong username, or wishes to change access methods, they may specify the {\tt --reset-authentication} option, which will delete the users file and allow the user to reenter their usernames. 

If errors occur during the checkout process, \rcltool\ stores the name of 
the component that caused the error, and prints out a list of all components that had errors before exiting. In addition any error will be logged, 
including the exact command that was called, and the error that was returned by the checkout tool. \rcltool\ will also time the entire checkout/update process and print the total time elapsed before exiting.

Multiple component lists may be specified together, in which case \rcltool\ will concatenate the lists and process them as one. The component list may also be specified as an URL, which \rcltool\ will download and then process normally. This further simplifies the code assembly process, as the user must only download \rcltool\ to initiate the assembly. In addition, the anonymous checkout process is shortened by performing a \emph{shallow checkout} of git repositories. As a distributed versioning system, cloning a git repository requires one to clone the entire repository, along with the full history of the repository. Over time, this history accumulates, and can consume a large amount of disk space. A \emph{shallow checkout} of a git repository only clones the most recent changeset, thereby reducing (sometimes greatly) the size of the resulting local copy, for example the Carpet repository can be reduced from 115MB to 76MB by performing a shallow checkout. 

\rcltool\ was written to be very modular, and it can easily be extended to include other versioning tools. All of the tools are handled by 
their own subroutine, and are pointed to by a single hash, which \rcltool\ compares with the \texttt{!TYPE} directive in each component. 
To add new functionality, one would only have to write a subroutine for the new tool, and add an entry to the \texttt{checkout\_types} hash.

\section{Example: Einstein Toolkit}
\label{examples}
\label{einstein}

The Einstein Toolkit~\cite{einsteintoolkitweb} is a collection of software components and tools for simulating and analyzing 
general relativistic astrophysical systems. Such systems include gravitational wave space-times, collisions of compact objects 
such as black holes or neutron stars, accretion onto compact objects, supernovae core collapse and gamma-ray bursts. Different 
research teams typically  use the Einstein Toolkit as the basis of their group codes where they supplement the toolkit with additional modules 
for initial data, evolution, analysis etc. 

The Einstein Toolkit uses a distributed development model where its software modules are either developed, distributed and supported by the core maintainers team, or by individual groups. Where modules are provided by external groups, the Einstein Toolkit maintainers provide quality control for modules for inclusion in the toolkit and coordinate support and releases. While the core of the toolkit is a set of Cactus thorns (distributed from different repositories), the toolkit also contains example parameter files, documentation, and tools for visualization, debugging, and simulation deployment. 

The component list (\texttt{einsteintoolkit.th}\footnote{https://svn.einsteintoolkit.org/manifest/einsteintoolkit.th}) for the Einstein Toolkit uses the CRL for distribution of its currently 130 different software components. All the components of the Einstein Toolkit are available by anonymous authentication as well as private authentication for the toolkit developers. 

The toolkit currently is distributed using SVN (Cactus Computational Toolkit, core Einstein Toolkit, parameter files, Simulation Factory), git (Carpet AMR driver), and CVS (components at CCT). A 
sample from the CRL file for the Einstein Toolkit is shown in Figure~\ref{et}.

\begin{figure}
{ \tt \small
\begin{Verbatim}[frame=single, framerule=0.3mm]
!CRL_VERSION = 1.0

!DEFINE ROOT = Cactus
!DEFINE ARR  = $ROOT/arrangements

# Cactus thorns
!TARGET   = $ARR
!TYPE     = svn
!AUTH_URL = 
https://svn.cactuscode.org/arrangements/$1/$2/trunk
!URL      = 
http://svn.cactuscode.org/arrangements/$1/$2/trunk
!CHECKOUT =
CactusArchive/ADM
CactusBase/Boundary
CactusBase/CartGrid3D
CactusBase/CoordBase
CactusBase/Fortran
CactusBase/IOASCII
CactusBase/IOBasic
CactusBase/IOUtil
CactusBase/InitBase
CactusBase/LocalInterp
CactusBase/LocalReduce
CactusBase/SymBase
CactusBase/Time

# McLachlan, the spacetime code
!TARGET   = $ARR
!TYPE     = git
!URL      = 
git://carpetcode.dyndns.org/McLachlan
!AUTH_URL = 
carpetgit@carpetcode.dyndns.org:McLachlan
!REPO_PATH= $2
!CHECKOUT = 
McLachlan/ML_BSSN
McLachlan/ML_BSSN_Helper
McLachlan/ML_BSSN_O2
McLachlan/ML_BSSN_O2_Helper
McLachlan/ML_ADMConstraints
McLachlan/ML_ADMQuantities
\end{Verbatim}
}
\caption{\label{et}Part of the CRL component list for the Einstein Toolkit.}
\end{figure}

\begin{figure*}[t]
\centerline{\includegraphics[width=\linewidth]{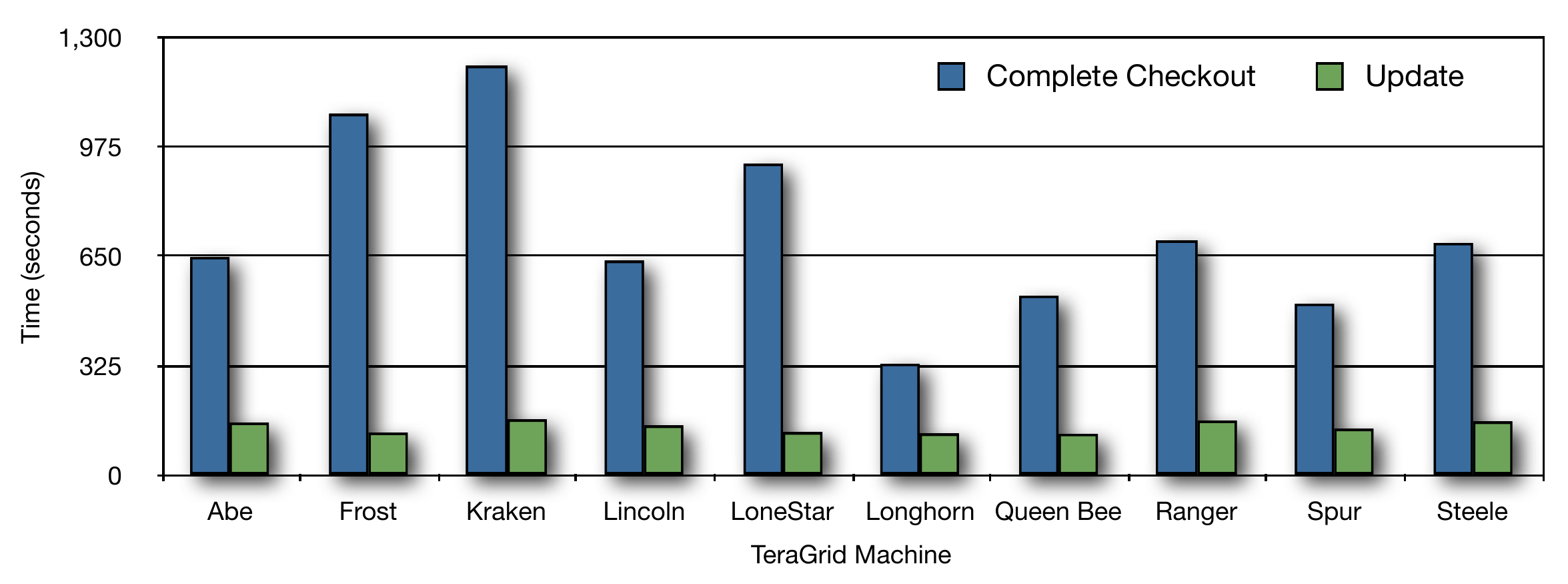}}
\caption{\label{results} Time taken for a complete checkout, and update, of the Einstein Toolkit with the GetComponents tool on different resources of the NSF TeraGrid.}
\end{figure*}

The \rcltool\ tool was tested using the Einstein Toolkit component on the resources of the NSF TeraGrid and the built-in timing mechanism was used to illustrate the time needed for both 
checking out and updating the full list of components (Figure~\ref{results}). While the testing was mostly successful, there were some issues. Notably, the Frost supercomputer at NCAR was using an outdated default version of Subversion, which was unable to process components using http or https protocols\footnote{The version of Subversion on Frost was updated to a working version just before submission of this paper.}. It was also difficult to establish a reliable connection to one of the CVS servers at CCT, so the tests did not include the two components from this repository, and they will likely be moved to Subversion in the near future.

\section{Future Work}
\label{future}

As illustrated in the results in Figure~\ref{results} assembling the Einstein Toolkit requires over 9 minutes on average. The time required for this could be reduced by introducing the concurrent checkout of different components. 

The CRL and \rcltool\ support the checkout and update of components. Source code versioning systems support many other options, including commits, tagging, and updates by date, version or tag. All of these features could be supported by extending the language to support distributed software development. For example, in the Einstein Toolkit consortium it could be helpful to remotely tag all the involved source code in releases of the Einstein Toolkit. One option that is currently being added is the ability to checkout or update the source code to a given date, to allow developers to more easily isolate the time and location at which a bug was introduced into the code base. With such a feature in \rcltool\ a wrapper script  could call \rcltool\ and run a regression suite repeatedly, to determine when software errors were introduced. 

Including provenance information is becoming a pressing challenge for scientific simulations. It is important that the code that produced published results can be easily reconstructed and rerun to reproduce or further analyze data. Currently, the Cactus Computational Toolkit contains a module, \texttt{Formaline}, which saves a copy of the complete simulation source code with the output data of a simulation. \rcltool\ could be extended to complement this by outputting a CRL file from a checkout that includes the information needed to recreate that checkout (depending on the particular system, this could for example be a version number or date associated with each component block or component). 

One issue that the CRL language and \rcltool\ do not address is how to construct a CRL file to solve a particular scientific problem. A future step will 
be to develop and implement a description language that describes the capability of components in such a way that users can query for components 
with a particular functionality. Further, the language should also be capable of describing dependencies between components. 

One possibility for identifying dependencies between components would be the use of an \texttt{!INCLUDE} directive, which would function similarly to the equivalent C directive. This extra directive would allow users to create an individual component list for each project, and use \texttt{!INCLUDE} to create a more logical structure for the framework, as opposed to listing every component in one large file or forcing users to always specify multiple files.

\section{Conclusions}
\label{conclusions}

This paper presented a language (CRL) that fully describes distribution mechanisms for software components for scientific codes. The \rcltool\ tool that implements   the CRL supports 
multiple source code versioning systems and other access methods, has the ability to checkout and update components and allows users to distribute and share component lists. 

The open source \rcltool\ tool should be of interest to collaborative teams of researchers with complex code bases using the NSF TeraGrid and other resources. 
\rcltool\ is now in production use with the Cactus Framework and the Einstein Toolkit.

\section{Distribution}
\label{distribution}

\rcltool\  is released under an open source license and is freely downloadable from \href{http://www.eseidel.org/projects/getcomponents/}{http://www.eseidel.org/projects/getcomponents/}.

\section*{Acknowledgments}

This work was supported by NSF \#0904015 (CIGR) \cite{ES-xirelweb},
NSF \#0725070 (Blue Waters), and NSF \#0721915 (Alpaca)
\cite{ES-Schnetter2007b, ES-alpacaweb}.  We used TeraGrid
resources at LONI, NCAR, NCSA, Purdue, and TACC under allocation
TG-MCA02N014.
The CRL and \rcltool\ implementation build on 
\texttt{GetCactus} which was originally developed by Gabrielle Allen and Tom Goodale.  We gratefully acknowledge suggestions provided by our colleagues 
in the Cactus and Einstein Toolkit projects, in particular Roland Haas for contributing darcs support to \rcltool. We thank Edward Seidel for providing comments on the paper. 

\bibliographystyle{amsplain-url}

\bibliography{RCL,publications-schnetter,references}
\end{document}